# Multi-scale variability in neuronal competition

Benjamin P Cohen[1], Carson C Chow[1], Shashaank Vattikuti[1,*]

1. Mathematical Biology Section, Laboratory of Biological Modeling, National Institutes of Diabetes and Digestive and Kidney Disease, National Institutes of Health, Bethesda, Maryland, United States of America

*E-mail: vattikutis@mail.nih.gov

## Abstract

We examine whether a single biophysical cortical circuit model can explain both spiking and perceptual variability. We consider perceptual rivalry, which provides a window into intrinsic neural processing since neural activity in some brain areas is correlated to the alternating perception rather than the constant ambiguous stimulus. The prevalent theory for spiking variability is a chaotic attractor called the balanced state; whereas, the source of perceptual variability is an open question. We present a dynamical model with a chaotic attractor that explains both spiking and perceptual variability and adheres to a broad set of strict experimental constraints. The model makes quantitative predictions for how both spiking and perceptual variability will change as the stimulus changes.

## Introduction

Variability is observed at multiple-scales in the brain. At the microscopic level, ion channels and synapses are subject to random effects of molecular discreteness[1,2]. Neocortical neurons fire stochastically and follow Poisson- or super-Poisson-like statistics[3–5]. At the cognitive level, variability is observed in behavior and perception. It is not clear how the variability at one scale is related to the variability at another scale. Variability at a small scale could induce variability at a larger scale or be averaged away and not be relevant. The nontriviality in the connection between microscopic and macroscopic variability played out previously in the attempt to explain spiking variability. While it was well known that ion channels and synapses are subject to small number biochemical variability, *in vitro* neuron spiking was found to be quite reliable when driven[6,7]. A resolution to this paradox invokes an attractor state with balanced excitatory and inhibitory synaptic inputs that yield a net input to neurons close to threshold so that fluctuations in the inputs drive spiking. Irregular spiking emerges robustly when the network settles into a chaotic attractor termed the *balanced state* [8,9]. In this case the variability is due to a deterministic albeit chaotic process. Here, we examine and quantify the relationship between spiking variability and perceptual variability.

Variability is ubiquitous in perception. It may serve a functional role for optimizing foraging[10], learning patterns such as songs[11], and for producing unpredictable trajectories while evading predators[12,13]. It may also help to arbitrate ambiguous circumstances such as that posed in the paradox of Buridan's ass who, equally hungry and thirsty, is placed precisely midway between a stack of hay and a pail of water and cannot decide. Perceptual variability can break this symmetry and release the ass from its fatal dilemma. The nature and source of perceptual variability is an open question. Although noise from the environment is important, perceptual variability is still observed when the stimulus conditions are controlled [2,14] and even when the eyes are paralyzed in a visual task (Leon Lack personal communication). Given that neuronal spiking is correlated with perception, spiking variability is a compelling etiology for the perceptual variability[5,15]. However, the precise mechanistic relationship between spiking and perceptual variability is unknown.

We focus on perceptual variability during neuronal competition and particularly perceptual rivalry. Neuronal competition is a ubiquitous property of the brain, playing a role in cognitive models of forced choice decision making[16,17], flanker-suppressor tasks[18], short term memory [19,20], and other computations [21,22]. Perceptual rivalry is a form of dynamic neuronal competition where the perception alternates between plausible interpretations given a fixed ambiguous stimulus and neural activity (in many brain regions) is correlated with the perception[23]. It is found in many visual contexts such as binocular rivalry, Necker cube, face-vase illusion, and motion-induced blindness, and also been reported for almost all sensory modalities[24]. The percept durations in rivalry obey a gamma-like distribution with coefficient of variation and skewness that is tightly constrained. This distribution is robust across many conditions, suggestive of intrinsic variability. It is found for both vision and audition[25], across species[26], and across a variety of visual stimulus conditions[27]. Despite the pervasiveness of the percept distribution there is no biophysical explanation for these statistics.

There is a long history of modeling neuronal competition with biophysically constrained cortical circuits [19,28,29]. For example, a circuit with lateral or mutual inhibition can exhibit *winner-take-all dynamics (WTA)* where a pool of neurons tuned to a percept suppresses the remaining pools [18,20,30]. With the inclusion of a fatigue mechanism, rivalrous alternations can arise from the WTA state[18,20,28,31]. However, rivalry is a challenge for quantitative modeling because of the many experimental constraints (see *Box 1*). Some models match the observed perceptual variability without adding noise but they fail to account for realistic spiking statistics (irregular, asynchronous spiking)[28]. Competition-like dynamics with variability have been demonstrated in deterministic balanced state networks but they have not been rigorously tested against perceptual constraints. An unstructured randomly connected network can produce alternating activity levels between two pools when receiving different fluctuating external inputs[32]. With structured connections and constant input, these models can exhibit temporary up-states or winnerless competition with balanced dynamics [14,33–35]. It has also been found that asymmetric activity levels in a balanced state can be achieved in mutual inhibition networks with a mechanism distinct from WTA[36]. A WTA network with mutual inhibition and high spiking variability has been invoked to explain choice probabilities in a perceptual decision-making task [37] but external noise contributed significantly to spiking variability. Thus, it remains to be seen whether balanced state and rivalry dynamics can coexist.

It is not clear that the balanced state can coexist with rivalry *prima facie*. Balanced state theory is predicated on a dynamic balance between excitation and inhibition. Rivalry strongly depends on imbalanced connections (e.g. mutual inhibition) between percept-encoding neuronal pools to produce a WTA state, and on a fatigue mechanism (e.g., spike-frequency adaption, synaptic depression), which is important for alternations. Neuronal adaptation has mixed effects. It can either aid irregularity by homogenizing synaptic inputs and facilitating a balanced state[38] or increase synchrony[39,40] and thus be antagonistic to commonly observed asynchrony. Finally, matching variability at one scale does not ensure matching at another scale without invoking additional mechanisms. For example, spiking variability may be too large or too small in magnitude or have no impact for perceptual variability. Thus, it is not *a priori* obvious how rivalry and all of its constraints can include biophysical spiking and whether the balanced state is a viable solution. Here we show that unstructured networks cannot explain rivalry but networks with structured mutual inhibition, adaptation, and network-induced biophysical spiking statistics can. We deploy balanced-state theory to show that this network breaks global balance; the dominant pool is balanced but the suppressed pool is not although it fires irregularly due to random input from the dominant pool. The model is also robust to connection architectures. In summary, we provide a self-consistent model for spiking and perceptual variability.

| **Box 1: Definitions and Experimental Constraints** |
|---|
| **Definitions** |
| Population – set of neurons in a synaptic class, i.e. a set of excitatory or inhibitory neurons |
| Pool – a network of excitatory and inhibitory neurons tuned to a percept |
| Dominant pool – actively spiking neuronal pool (determined by the excitatory population activity), which corresponds to dominant percept |
| Suppressed pool – weakly active or inactive neuronal pool corresponding to suppressed percept |
| Dominance duration – time duration dominant pool remains highly active |
| Report threshold – minimum reportable event duration |
| Drive – external feedforward current (mV/msec) |
| **Experimental Constraints** |
| **Perceptual constraints - Levelt's proposition[41]:** |
| 1. Mean dominance duration is on the order of seconds. |
| 2. Levelt's 4th proposition – mean percept duration decreases with increased drive |
| 3. Classical Levelt's 2nd proposition – starting from the point where the stimulus drive gives equal dominance durations across the pools (equi-dominance), then weakening drive to one pool increases <u>the opposite</u> pool's predominance |
| 4. Modified Levelt's 2nd proposition – from the equi-dominance point, strengthening drive to one pool increases <u>the same</u> pool's predominance |
| 5. Maximal alternation rate – when modifying drive to one pool, the rivalry alternation rate is fastest when both pools have equal dominance times[42] |
| **Variability constraints:** |
| Perceptual variability: |
| 1. Dominance-time durations follow a gamma distribution with coefficient of variation ($CV_D$) between 0.4 and 0.8[27] |
| 2. Skewness of dominance time distribution is twice the $CV_D$ |
| (Additional constraints for our analysis to ensure the distribution is from the dynamics and not posthoc filtering.) |
| 3. Dominance statistics are robust to chosen report threshold |
| 4. Mode of dominance time distribution is to the right of the report threshold |
| |
| Spiking variability: |
| 1. Mean spike-count correlations less than 0.3 [43] |
| 2. Spike count Fano Factor is between 0.9 and 2.0 [4] |
| 3. Coefficient of variation for inter-spike intervals is between 0.5[3] and 2[7] (target of 1.0 used for model fitting) |
| 4. Mean spike-rates are between 5 and 40 Hz during up-states, and below 10 Hz when suppressed [4] |

## Methods

### Model neurons

Neuronal dynamics were modelled as leaky integrate-and-fire neurons with adaptation that obey

$$\frac{dv_i}{dt} = f_i + s_i(t) - \frac{v_i}{\tau_m} - \gamma a_i(t), \qquad v_i < \theta \qquad (1)$$

$$v_i \to v_i - \theta, \qquad v_i = \theta \qquad (2)$$

$$\frac{ds_i}{dt} = -\frac{s_i(t)}{\tau_s} + \sum w_{ij}\delta(t - t^j_{spike}) \qquad (3)$$

$$\frac{da_i}{dt} = -\frac{a_i}{\tau_a} + \delta(t - t^i_{spike}) \qquad (4)$$

where $i, j$ are neuron indices, $v$ is neuron voltage, $f$ is feedforward current drive, $\tau_m$ is the membrane time constant, $w$ is the synaptic strength from neuron $j$ to $i$, $s$ is synapse strength, $\tau_s$ is the synaptic time constant, $a$ is an adaptation variable, $\gamma$ is the adaptation strength, $\tau_a$ is the adaptation time constant. We used Euler's method for simulations and at a given time step, any neuron that was above threshold was reset as its current voltage minus threshold. Spike times ($t_{spike}$) were interpolated. Computer code available at https://github.com/ShashaankV

### Network architectures

We studied three randomly connected cortical circuit architectures (see Figure 1): unstructured network, discrete mutual inhibition network, and structured continuum network. All network simulations had a total of 4,000 neurons. We checked to see that results did not qualitatively change with neuron number. The *unstructured network* was evenly split between excitatory and inhibitory populations. Each neuron received *k* randomly chosen connections of each neuron type (excitatory or inhibitory), with different synaptic strengths depending on the type of synapse. The *discrete mutual inhibition network* consisted of two unstructured-network pools, with 2,000 neurons forming each "percept pool". The two pools were linked by *k* randomly chosen long-range connections from excitatory to inhibitory neurons ($w_{ie_{long}}$). The *continuum network* consisted of 80% excitatory and 20% inhibitory neurons. Neurons were arranged evenly on a ring and synaptic strengths decayed as a function of the distance between neurons, obeying von Mises distributions (see *Supplementary Material: Synaptic coupling in continuum model*). Connections were randomly deleted with survival probability *p*.

### Drive

We manipulated feedforward drive to neurons in several ways. For the unstructured network we examined two cases: homogenous and heterogeneous drives. In the homogeneous case, excitatory neurons received the same non-fluctuating drive that was slightly higher than the inhibitory neurons. For the heterogeneous case, the network was divided into two excitatory pools each receiving an independent fluctuating drive, modelled as Ornstein-Uhlenbeck stochastic processes. All structured network simulations received non-fluctuating drive to only the excitatory neurons. For the discrete mutual inhibition model, all excitatory neurons received a stimulus drive. For the continuum model, only a subset of excitatory neurons received this input. Levelt's propositions were investigated by changing drive strengths as detailed in Figures and computer code.

**Measures**

Dominance durations were estimated by converting spike rates across pools into a *percept state variable P*, where $u_A$ and $u_B$ were the sum of spikes across excitatory neurons of each pool for 50 millisecond time windows. In each window, the percept state was assigned to $(u_A - u_B)/(u_A + u_B)$, which results in a number between -1 and 1. We divide this domain into even thirds to classify the percept state: percept A (P > 1/3), percept B (P < -1/3), and neither percept A or B (-1/3 < P < 1/3). A dominance duration was measured as the interval between state changes. Spike count Fano factor and spike count correlations were estimated by sampling 100 millisecond windows for each neuron and then calculating statistics from this. The interspike-interval coefficient of variation ($CV_{ISI}$) was estimated for each neuron across simulation blocks. We isolated dominant vs suppressed states and calculated statistics for each state.

**Figure 1**

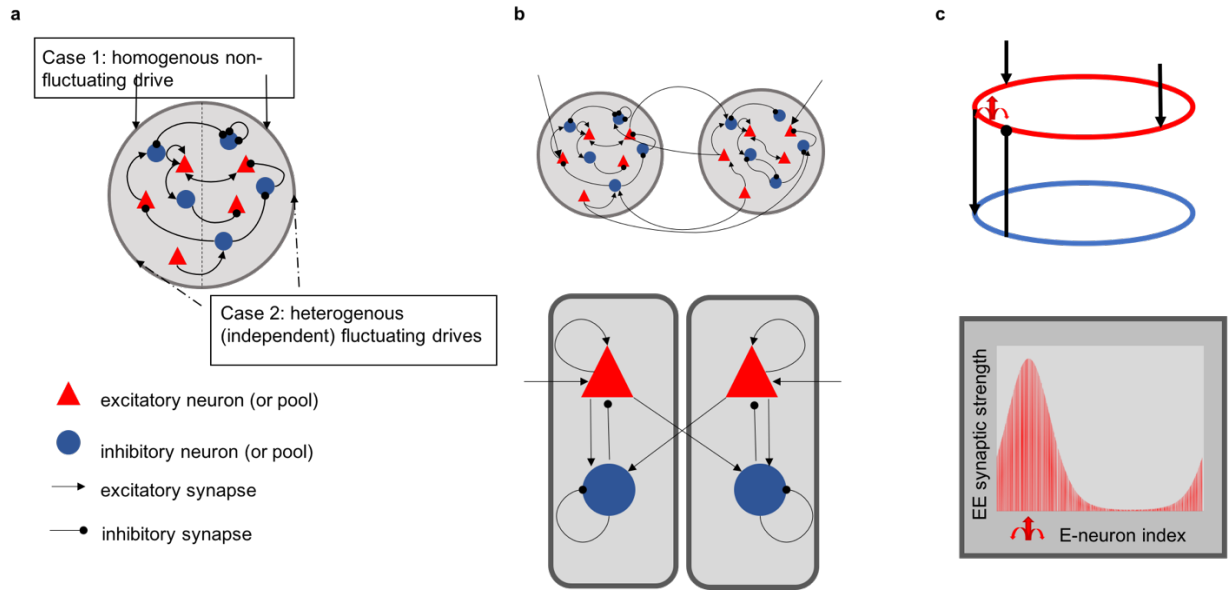

**Figure 1. Three network architectures. a)** Unstructured network receives either a homogenous non-fluctuating drive (Case 1) or each half of the network receives independent stochastic (fluctuating) drive (heterogeneous drive, Case 2). **b) (top)** Discrete mutual-inhibition network consists of two pools of excitatory and inhibitory neurons tuned to a specific percept. Each pool is an unstructured network as in **a** and receives non-fluctuating drive to excitatory neurons only. **(bottom)** Mutual inhibition is modelled as long-range connections from excitatory neurons in one pool to inhibitory neurons in the other. **c) (top)** Continuum network consists of neurons deposited along two rings, excitatory and inhibitory, with spatially structured coupling between all neuron types. Two sets of excitatory neurons on opposite sides of the ring (representing different percept tuning) receive non-fluctuating feedforward drive. **(bottom)** Example of the spatial profile for excitatory-to-excitatory synaptic strength from single presynaptic neuron. Strength is periodic and maximal at the presynaptic neuron (indicated by red arrow).

## Results

*Unstructured network does not capture rivalry*

We evaluated competition dynamics in an unstructured network with balanced-state dynamics and tested if it could match the empirical constraints of rivalry in *Box 1*. Homogenous drive resulted in biophysical spiking and statistically homogenous irregular activity as predicted by balanced-state theory (Figure 2a) [8,9]. The addition of a fatigue mechanism did not rescue the model (see *Supplementary Material: Adaptation effect in the unstructured network*). If subsets of neurons receive heterogeneous drive (Figure 2b) then competition between two pools can emerge[32]. This resulted in epochs where one pool had a higher spiking rate than the other (Figure 2c). However, this network fails to capture Levelt's 4th proposition. Instead of percept durations decreasing with increased drive, they increased (Figure 2d). In addition, the dominance duration distribution was not stable to changes in the report threshold (Figure 2e). As shown in Figure 2f, the competitive dynamics in the unstructured network mirrors the external drive [32]. The tight correlation (r = 0.94) between the drive and the network activity demonstrates that alternations result from the network activity tracking and amplifying differences in the feedforward drive. As noted before[44] and shown below, the heterogenous drive breaks global balance. The unstructured network does not generate rivalry dynamics itself. However, it could satisfy all empirical constraints if it received inputs from a rivalry source that satisfied the perceptual constraints if not the spiking constraints. However, this leaves the original source of the rivalry dynamics unexplained.

Figure 2

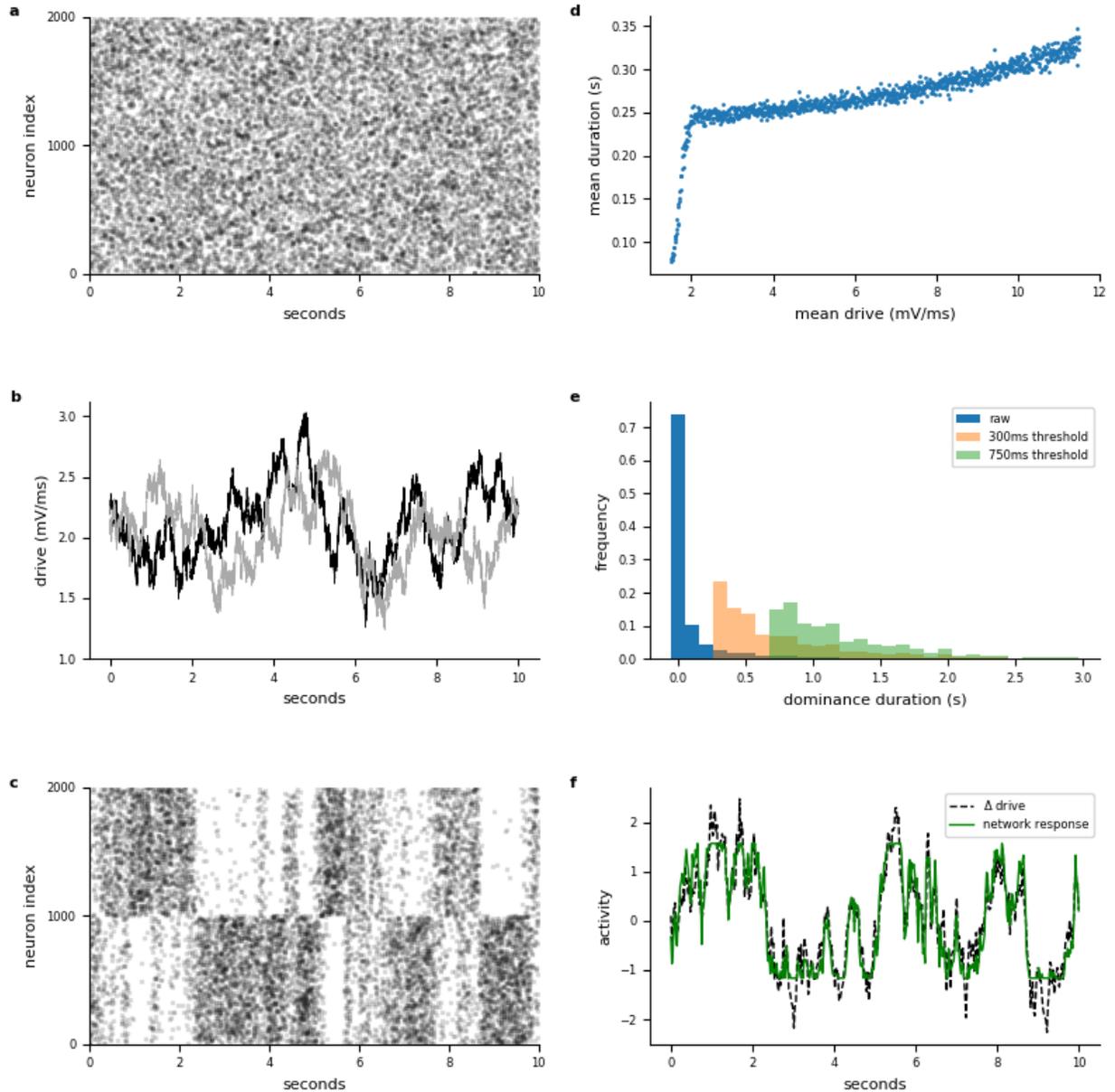

**Figure 2. Unstructured network does not comply with rivalry constraints. a) Case 1: Homogenous drive raster** (excitatory neuron spiking across time). **Case2: Heterogeneous drive (b-f) b)** heterogeneous drive consists of independent stochastic processes for subsets of neurons within the unstructured network. **c)** Excitatory neuron raster showing alternating activity levels between two pools in response to the heterogeneous input in **b**. **d)** Inverse Levelt's 4th proposition: percept durations increase with drive strength. **e)** Dominance time distribution of random network (as in **c**). Dominance duration coefficient of variation ($CV_D$) depends on the *report threshold* (see *Box1: Definitions*). For example, $CV_D$ are 1.9, 0.61, and 0.34 for 0, 300ms, and 750ms report thresholds respectively. **f)** Percept state variable (z-scored) reflects differences in the drives in **b**, supporting that the dominance duration statistics are derivatives of the drive fluctuations.

*Structured networks can satisfy all constraints of rivalry*

We evaluated two structured cortical circuit network architectures. One was a discrete mutual inhibition network that was previously used to model rivalry, flanker-suppressor tasks, and normalization[18] (Figure 1b). The second was a modification of a continuum model used previously for rivalry and other tasks [19,28] (Figure 1c). None of these previous models had been shown to satisfy all of the empirical constraints in *Box 1*. We numerically scanned parameter space in both models to find regions where the rivalry constraints were satisfied. The networks all received constant, non-fluctuating drive and were completely deterministic. Figures 3 and 4 show results for single representatives of each network architecture (discrete and continuum) that matched the constraints. A summary of matched perceptual constraints is shown in Figure 3, where points are multiple realizations of a single model. A summary of variability results is shown in Figure 4 for a single realization.

To test Levelt's propositions we measured the response to changes in drive strength (see Figure 3). Levelt's 4th proposition describes the effects of symmetric drive. As shown in Figure 3a, the dominance times decreased with strength in both models, in keeping with Levelt's 4th proposition. Figure 3b shows that a plot of SD vs mean of dominance duration is well fit by a regression line with the empirically observed slope $CV_D$ [27]. However, when $CV_D$ is computed independently for each drive strength, it shows a small but significant decreasing trend with increasing drive strength while staying within the empirical range (Figure 3c). This trend could serve as a falsifiable prediction for the model. In response to asymmetric drive strengths (Figures 3d-e), both models matched the characteristic dominance duration profiles seen in Moreno-Bote et al.[42] and produced the characteristic 'X' shape of the classical and modified Levelt's 2nd proposition. The alternation rate is maximal when both populations have the same dominance duration. Thus, the models matched all of the Levelt's proposition constraints.

**Figure 3**

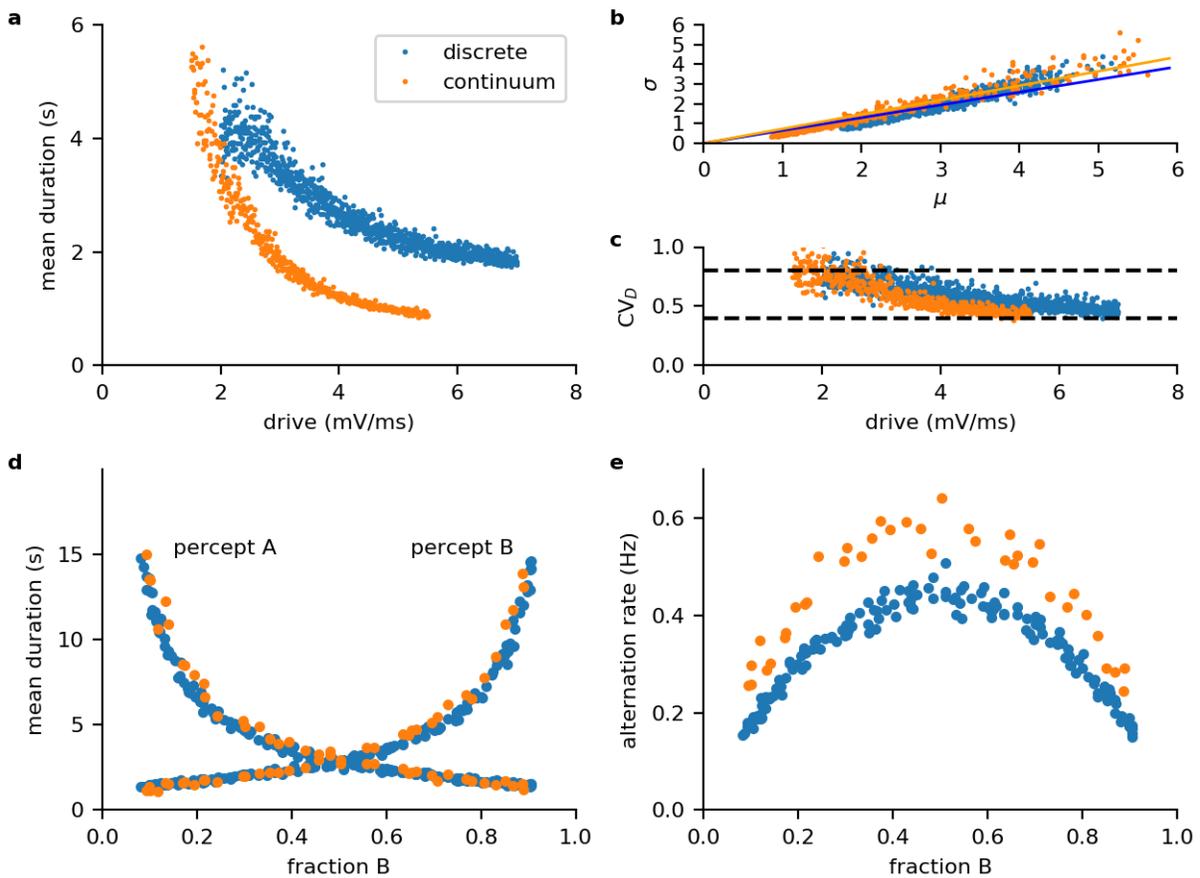

**Figure 3. Psychophysics reproduced by discrete mutual inhibition and continuum networks. a)** Networks match Levelt's 4th proposition; dominance duration decreases with drive strength. **b)** SD ($\sigma$) vs mean ($\mu$) dominance duration is well fit by regression line with slope $CV_D$ matching Cao et al.[27] (discrete slope=0.65, P-value <$10^{-314}$; continuum slope=0.73, P-value <$10^{-314}$). **c)** $CV_D$ computed at each drive strength stays within the experimentally observed range (dashed lines) across changes in drive but has a significant trend (P-value<$10^{-159}$). **d)** Networks match Levelt's 2nd and modified 2nd propositions (compare to figure 6 in Moreno-Bote et al.[42]) **e)** Overall alternation rate decays symmetrically from equal dominance (compare to figure 8 in Moreno-Bote et al.[42]).

Spiking and perceptual variability constraints could be satisfied by both models while still satisfying Levelt's propositions. Dominance durations were irregular (see Figure 4a) and showed gamma-like statistics (Figure 4b-c). The distributions matched the shape parameter from experiments on real subjects[45] (indicated by dashed lines) automatically without imposing any additional constraints. In addition to matching percept variability, the networks exhibited biophysical spiking (Figure 4d-g). The model exhibited irregular (Figure 4d,f) and asynchronous (Figure 4e,g) spiking. The average inter-spike-interval coefficient of variation $CV_{ISI}$ was 1.22 during dominant states and 1.19 during the suppressed states. The average spike-to-spike correlations $r_{sc}$ was 0.02 during dominant states, and 0.06 during suppressed states. All model spiking statistics fell well within the empirically reported ranges (indicated by dashed lines). As shown in Figure 4h, we found a relationship between drive and spiking variability but in the opposite direction of $CV_D$ as noted above. Increasing drive strength decreased perceptual variability but increased spiking variability, when Levelt's 4th proposition is obeyed. This forms an empirical prediction and test of the model during realistic stimulus conditions.

Overall it was easier to match experimental constraints in the discrete versus the continuum model. For the discrete model, rivalry dynamics were found by starting with intra-pool parameters that led to biophysical spiking in each pool, then adjusting mutual inhibition strength until WTA appears. We then adjusted the adaptation strength and time constant to achieve rivalrous alternations. Since adaptation can affect spiking variability, parameters often had to be readjusted to recover biophysical spiking. Once both spiking and perceptual variability converged to the empirical constraints then, remarkably, rivalry dynamics satisfying all empirical constraints naturally emerged. For the continuum model, we used a random sampling approach to find conditions that matched all constraints since the effect of the parameters on local and global dynamics were not easily untangled. In numerical experiments, we found that in the continuum model, parameters for WTA were about tenfold more difficult to locate than for biophysical spiking, and that the combination of the two was rare but not overly difficult to find (*Supplementary Material: Continuum model parameter search*).

Figure 4

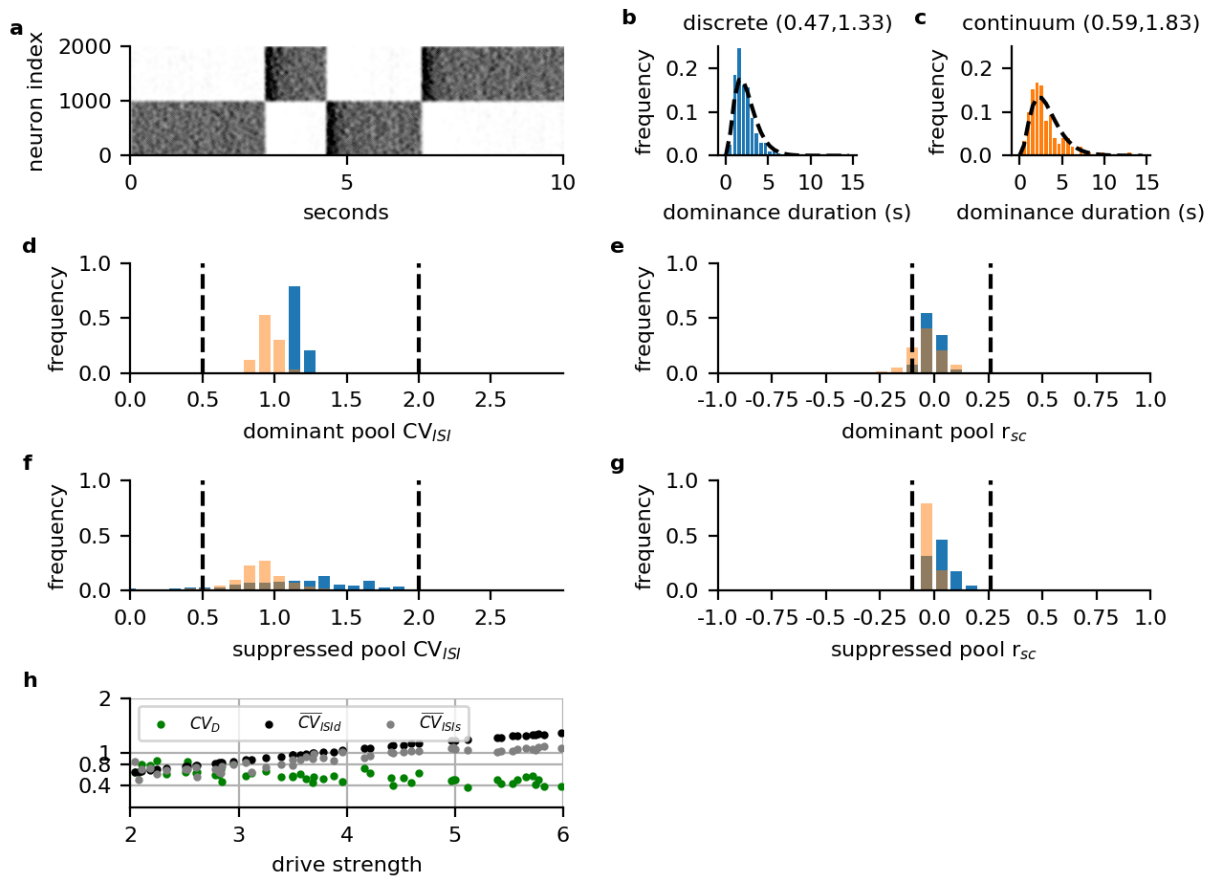

**Figure 4. Biophysical variability at multiple scales in discrete mutual inhibition (blue) and continuum (orange) networks. (a-g)** Results from single realization of example systems (symmetric drive=5). **a)** Excitatory neuron raster of discrete mutual inhibition network during rivalry showing stochastic dominance durations and spiking. **b-c)** Gamma-like distribution of dominance durations with 300ms report threshold (see *Box 1*). $CV_D$ and skewness, respectively, are given in parenthesis. Dashed lines in **b-c** are the empirical gamma distribution shape parameter from Robertson et al. 2013 during binocular rivalry[45] using the scale parameter from our simulations. **d-e)** Dominant-pool excitatory neuron spiking statistics. Distributions from per-neuron statistics. Distribution of interspike-interval coefficient of variation ($CV_{ISI}$) and spike-count correlation ($r_{sc}$) reflect an irregular and asynchronous spiking. **f-g)** Suppressed-pool excitatory neuron spiking statistics. Mean $CV_{ISI}$ is near one but has wider variance than dominant state. The mean $r_{sc}$ is near zero but skewed towards positive correlations compared to the dominant state. Dashed lines in **d-g** are empirically observed bounds (see *Box 1*). **(h)** Spiking and perceptual variability as a function of drive strength obeying Levelt's 4th proposition. Plot of $CV_D$ and the average $CV_{ISI}$ across neurons in the dominant ($\overline{CV}_{ISId}$) or suppressed ($\overline{CV}_{ISIs}$) states. There was a significant linear trend for all measures (maximum P-value<$10^{-8}$). Measures remained in the empirical constraints as indicated by y-ticks.

*Two-pool balanced-state theory and mixed states*

We examined more closely the nature of the network dynamics using balanced state theory [8,9]. We focused on the WTA state; our analysis can be applied to the rivalry state as long as the adaptation time constant is slow compared to the spiking frequency. Consider a network of two pools, each with an excitatory and an inhibitory population. Let **W** be the matrix of the average connection strength between all populations, **r** be the vector of mean spiking rates of each population in both pools and **f** be the vector of external drive to each population in both pools. Balanced state theory proposes that in the large system limit, the net mean input to each population is zero. In matrix notation, this is expressed as

$$\mathbf{Wr} + \mathbf{f} = 0 \qquad (1)$$

Consider first the unstructured random network where the external inputs differ between two pools of neurons that are otherwise identical. As noted previously[44], we immediately see that a global balanced state solution cannot exist since the rows of **W** are symmetric between the two pools while those of the external input vector **f** are not and thus a solution for **r** to equation (1) cannot be found since **f** is not in the column space of **W**. However, a solution is possible if we break the symmetry between the intra- and inter-pool connections. In our structured discrete network, each individual pool is identically connected but the two pools are connected only through excitatory to inhibitory connections. This sparse matrix allows symmetry between the pool rates to be broken and a solution for **r** is possible. Details of all calculations are given in the *Supplementary Material: Balanced-state theory*.

We quantitatively compared the predicted rates from balanced-state theory to the spiking network by solving (1) for **r**. Since our simulation is for a finite number of integrate-and-fire neurons and it is not clear what the exact threshold should be, we allow the thresholds for excitatory and inhibitory neurons to be free parameters that we adjust to match theory with simulations. Figure 5 shows a comparison over a range of mutual inhibition strength, $w_{ielong}$. For weak mutual inhibition, both pools are active and the network spiking rate is consistent with the theoretical prediction for a two-pool balanced state. However, at a critical value of $w_{ielong}$, there is a singularity in the theory and the two-pool balanced-state theory breaks down. The critical $w_{ielong}$ is consistent with the transition between a state where both pools are active to the WTA state in the simulation. During WTA, the dominant pool receives virtually no input from the opposing circuit and thus reduces to a local, classical one-pool balanced state network. The suppressed pool is not in a balanced state because excitation cannot balance the inhibitory input from the dominant pool, which is the opposite case of Ref 44. However, the network still fires irregularly and asynchronously because it is driven by the balanced chaotically firing dominant pool.

Also consistent with balanced-state theory, the mean input is largely invariant relative to the change in mean spike-rate when pools are in the balanced state. By contrast, inputs become variable and decrease with mutual inhibition strength when they are not balanced (see Figure 5e-f) (also see *Supplementary Material: Inputs in mixed balanced network*).

**Figure 5**

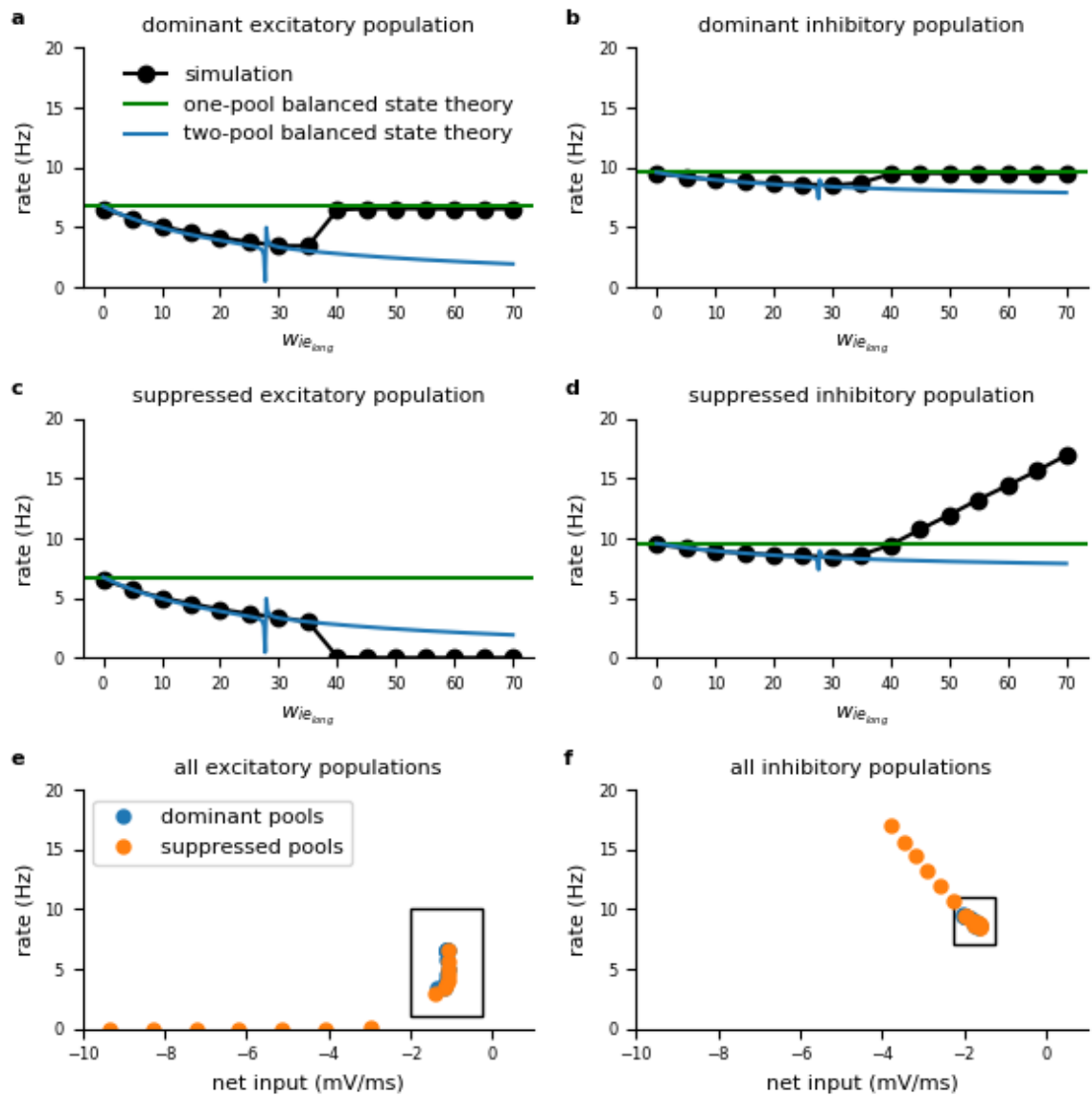

**Figure 5. Balanced and imbalanced states in theory and simulation during symmetric activity and WTA. a-b)** The dominant pool spiking rates obey the two-pool balanced state theory before the WTA transition. After the transition, they obey the one-pool theory. The discontinuity in the two-pool prediction is a singularity where the theory breaks down (same for **a-d**). For the simulation, there is a change in the network state soon after this point. **c-d)** The suppressed pool obeys the two-pool theory prior to WTA. It deviates from balance after the transition. The excitatory population is quiescent, whilst the inhibitory population receives balanced excitatory input from the dominant pool and fires rapidly. **e-f)** While in the balanced state, the net inputs (synaptic and external drive) to excitatory and inhibitory neurons form clusters near threshold (indicated by boxes) where inputs balance.

**Discussion**

We asked if both psychophysical responses (specifically rivalry) and the balanced state can coexist in a single canonical circuit model, and thus provide a self-consistent model for spiking and perceptual variability. We found that the answer is yes and no. Indeed, spiking and perceptual variability (and other psychophysical phenomena) are explained by the same deterministic, chaotic attractor system. However, further investigation of the balanced state theory reveals a deviation from pure balance. The two-pool balanced state theory we used is applicable to WTA dynamics and cases where both percept-pools are active such as in normalization[18,46]. When we analyzed the mechanism for biophysical spiking and WTA, we found that the dynamics are due to a mixture of balanced and imbalanced states within the network. This gives an interesting twist to current notions of spiking in the brain. Irregular, asynchronous spiking may imply that the brain is either in a balanced state or it is being driven by a balanced state.

Our findings also augment our understanding of rivalry. Rivalry is an alternation in percept states that depends on neuronal fatigue, such as spike-frequency adaptation or synaptic depression. During a dominance epoch, the winning or dominant pool is above a neuronal threshold and the suppressed pool is subthreshold. There are two dynamical possibilities for how alternating dominance could occur. The first is called *release*, where a dominance switch occurs when neurons in the dominant pool fatigue to the point of falling below the spiking threshold and stop spiking. The second is *escape*, where neurons in the suppressed pool recover from fatigue and overcome the suppressing inhibition and thereby become dominant[47]. Levelt's propositions require the escape mechanism[31]. This necessity is clearly seen in Levelt's 4th proposition which states that dominance duration increases with decreasing drive strength. A release mechanism would predict the opposite since decreasing drive would mean that a fatiguing neuron would drop below threshold faster and thus decrease dominance duration. However, under escape, a decreased drive would prolong the time of a suppressed neuron to recover from fatigue and spike and thus increase the dominance duration. The dominance duration is given by the time to escape, which is governed by the time constant of the fatigue mechanism (e.g., adaptation). However, the precise moment of escape is determined by the net input to the neuron as a function of time. If the neuron input is subject to uncertainty or noise then the time to escape can be stochastic and if the approach to threshold is shallow then even a small amount of noise can lead to a large amount of dominance duration variability. We show that self-induced chaotic variability of the balanced state in the dominant pool is sufficient to act as the noise source in the suppressed pool to explain perceptual rivalry. This implies a crucial role for cross-multiplicative noise (through synaptic inputs) in stochastic versions of rivalry and canonical circuit models.

Further examination of our model, including different neuron models and architectures, is warranted to discover the extent and robustness of our findings. Our analyses were also restricted to the two-pool competition case. This represents the minimal competition model that explains many general rivalry effects. However, specific effects such as mixed perception in binocular rivalry may require three or more pools. The model can be expanded for these cases. We propose a refinement to a prior hypothesis for the invariance of perceptual variability across stimulus conditions. Cao et al.[27] noted that perceptual variability ($CV_D$) was around 0.6 across many competition conditions from binocular rivalry to motion-induced rivalry. This led them to hypothesize an invariance for rivalry variability. However, we found that increased drive strength can decrease the $CV_D$; though, some parameter cases were more stable.

This suggests that there may be finer structure within the experimental noise observed in Cao et al.. Similarly, we observed a spiking variability effect due to drive (when controlling for dominant or suppressed state), but in the opposite direction of perceptual variability. Notably this is within the stimulus range of behavioral experiments and forms an additional test of the model. It is paramount to test these model predictions in experiments to further refine the model. The apparent anti-correlation is interesting to note and raises question regarding the relationship between spiking and perceptual variability. Further theoretical work is warranted to investigate this relationship between drive, scales, and across other circuit parameters.

Does this model have utility and implications beyond rivalry? Rivalry stands out among perceptual phenomena since concrete changes in perception occur despite a static stimulus, thus providing insight into internal workings of the brain. It has even been considered a tool for investigating neural correlates of consciousness[48]. It is not surprising then that rivalry is one of the target behaviors of a developing canonical cortical circuit theory of cognition[20]. Similar circuit architectures are used to explain "cognitive primitives" such as short-term memory and decision-making and have been used to interpret differences in cognitive measures among clinical, psychiatric cohorts[49–52]. A major challenge of psychiatric models is how to scale from the molecular perturbations underlying mental illness to complex psychopathology. Here we present a self-consistent theory for spiking and perceptual variability that bridges two important levels. However, this raises the question of the role of molecular variability, which exists in many forms[2]. It does not seem to be sufficient to explain spiking variability[9,32] and, in light of our findings, it is unnecessary for a major form of perceptual variability. However, since the balanced state keeps neurons near threshold due to a chaotic attractor then small perturbations due to molecular noise could still have large effects. This will be interesting to study in the future. We propose that the influence of molecular variability is already contained within a set of hyperparameters of the cortical circuit governing the distribution of time constants and synaptic strengths. Theories that tie together biology at multiple scales will be useful for making psychiatric and cognitive problems tractable. Immediately one application is in addressing the *excitation-inhibition (EI) imbalance hypothesis*[20,52,53] in mental illness, a pervasive but ill-defined hypothesis in clinical research. The model here is based on two important EI relationships, the balanced state and mutual inhibition, and thus is a good candidate for exploring these questions. EI can be manipulated in the model to make clinical research predictions at multiple scales of behavior. The model can also be used as middle ground to map clinically-associated molecular factors to the circuit parameters to make further predictions.

**Author Contributions**

BPC performed the majority of computational experiments and wrote the computer code; SV performed additional experiments, analyses, and edited code; CCC and BPC developed the mathematical theory; SV, CCC, and BPC wrote the paper. This work was supported by the Intramural Research Program of the NIH, NIDDK. The funders had no role in study design, data collection and analysis, decision to publish, or preparation of the manuscript.

**Competing Interests**

The authors declare no competing interests.

**Multi-scale variability in neuronal competition: Supplementary Material**

Benjamin P Cohen[1], Carson C Chow[1], Shashaank Vattikuti[1,*]

1. Mathematical Biology Section, Laboratory of Biological Modeling, National Institutes of Diabetes and Digestive and Kidney Disease, National Institutes of Health, Bethesda, Maryland, United States of America

*E-mail: vattikutis@mail.nih.gov

Contents

**1. Synaptic coupling in continuum model**

The continuum network consisted of 80% excitatory and 20% inhibitory neurons. Neurons were arranged evenly on a ring and synaptic strengths decayed as a function of the distance between neurons (in radians), obeying von Mises distributions:

$$w_{ij}^* = \frac{A^* \exp(\kappa^* \cos(\theta_i - \theta_j))}{pN^* 2\pi I_o(\kappa^*)} \text{ with probability } p, \tag{1}$$

otherwise $w_{ij}^*=0$

where $w_{ij}^*$ is the synaptic weight for synapse class * (e.g., excitatory-to-excitatory, inhibitory-to-excitatory) from presynaptic neuron *j* to postsynaptic neuron *i*. The parameter *A* governs the amplitude, $\kappa^*$ governs the decay rate of strength over distance, *N* are the number of postsynaptic neurons, *p* the probability of a synapse, $\theta$ the neuron location in radians, and $I_o$ is the Bessel function of order 0.

## 2. Continuum model parameter search

In the continuum model, we randomly sampled synaptic coupling parameters without adaptation in a million networks each with 2000 excitatory neurons and 500 inhibitory neurons. Out of 1,000,000 simulations, 713,608 completed successfully. Failed simulations typically had excessive recurrent excitation which caused an inordinate number of spikes so that the simulation did not run to completion in the allotted time. Equations were solved using the forward Euler method, which we implemented using Julia version 0.4.5.

We classified networks according to the following (nonexclusive) dynamical properties: 1) *Winner-take-all* (WTA) meaning that 90% of the total number of spikes were from one population, 2) *asynchronous* meaning that mean spike-count-correlations within a population were less than 0.1, and 3) *irregular* meaning that the spike count Fano factor was between 0.7 and 2.5. 46.2% of all models exhibited asynchronous and irregular spiking characteristics, whereas only 3.9% of models exhibited winner-take-all behavior. 26.9% of winner-take-all models had asynchronous irregular spiking. Networks with both irregular and asynchronous spiking were deemed *biophysical*.

We sampled 2,000 models randomly from the collection exhibiting WTA and biophysical spiking, and scanned adaptation parameters in each one. To select reliable models, we isolated models with data for at least 20 simulations at different adaptation strengths. A small number of simulations would fail, leaving 1996 reliable models. Of these, 1949 exhibited rivalry with mean dominance time longer than 1 second in one population for at least one set of adaptation parameters. This showed that the majority of models with WTA produce rivalry when adaptation is introduced.

The effect of adaptation on dominance time and spiking statistics was slightly different for each model. The coefficient of variation of dominance times, mean dominance time, and spiking statistics can be affected simultaneously, so a successful model has adaptation parameters that align all three measures in realistic regimes. A parameter Venn diagram representation of models achieving success is shown in Figure S1.

**Figure S1**

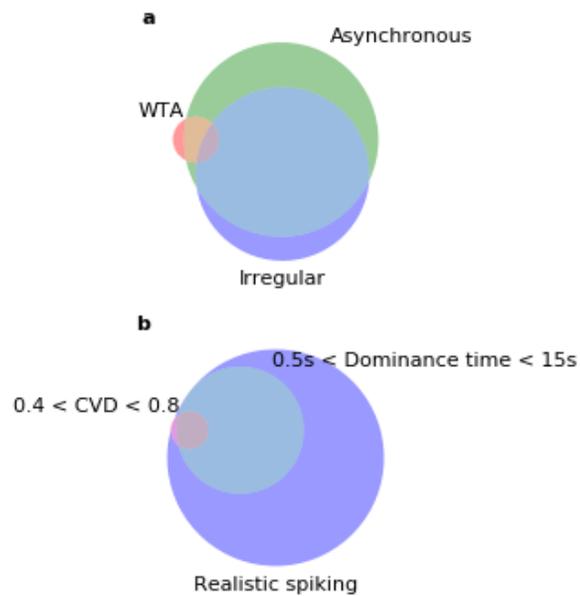

**Figure S1: Fractions of parameter space in continuum model satisfying various constraints. a)** Arrangement of asynchronous, irregular, and WTA regimes in the 713, 608 models for which we had simulation data. **b)** A single model complying with all the constraints in (A) was selected and scanned for adaptation parameters. The arrangement of regimes with dominance times on the order of seconds, $CV_D$ in the experimental range, and realistic spiking is shown.

## 3. Inputs in mixed balanced network

A balanced state has been used to explain biophysical spiking statistics. An intriguing question is whether this is the case for the WTA model with biophysical spiking. We found that this is not the case. Being balanced means that the mean net input does not change when the individual components are changed. To examine this (see Figure S2), we compared inputs to the pools when they had equal activity (middle row) versus in the winner-take-all state (top: looser and bottom: winner rows). During WTA, the mean input to the winning (dominant) pool matches the equal activity state but the loosing pool mean is shifted downward. Thus, rather than being balanced by recurrent excitation, input in this pool decreases and firing rate drops. This suggests that within the same network, one pool is balanced and another is imbalanced simultaneously.

**Figure S2**

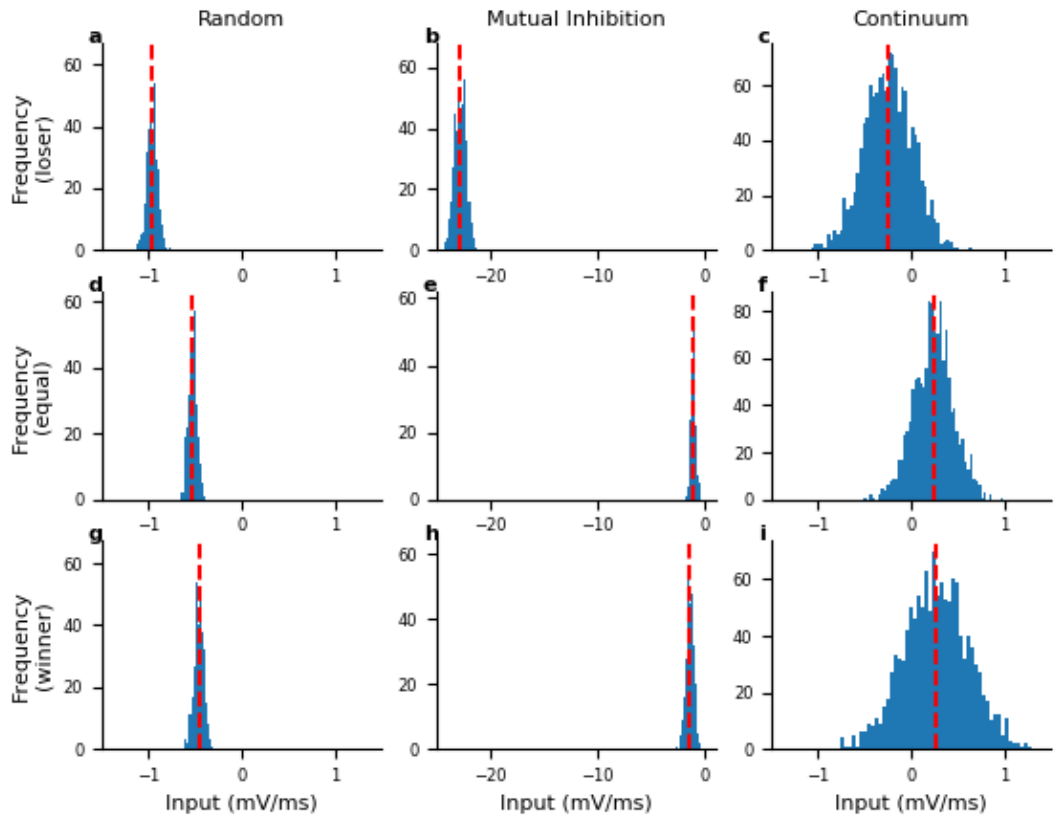

**Figure S2: Net inputs to three models during equal and winner-take-all states. a,d,g)** Unstructured network with homogenous drive giving equal activity. Net input in this case is shown in **d**. Network given heterogeneous drive causing elevated activity in one pool (winner) and lower activity in the other (looser). Net input to winner and loser pools shown in **a** and **g**, respectively. **b,e,h)** Discrete mutual inhibition model. Low mutual inhibition and homogenous non-fluctuating drive results in equal activity between pools. Net input for this case is shown in **e**. Increasing mutual inhibition with similar drive results in asymmetric activity winner and loser pools. Net input to winner and loser pools shown in **b** and **h**, respectively. **c,f,i)** Continuum model. Broad recurrent excitation and homogenous non-fluctuating drive results in equal activity between pools. Net input for this case is shown in **f**. Narrowing recurrent excitation with similar drive results in asymmetric activity winner and loser pools. Net input to winner and loser pools shown in **c** and **i**, respectively.

## 4. Balanced-state theory

Consider first a random network of a single pool with an excitatory and an inhibitory population. We express the total inputs to neurons in terms of the spiking rates and coupling strengths of each population. Balanced-state theory argues that the excitatory and inhibitory inputs to neurons must balance near threshold in the limit as N goes to infinity to avoid a blowup of the spiking rates. This leads to a system of two equations – one for the input to the excitatory population, and one for the input to the inhibitory population:

$$I_e = w_{ee}r_e - w_{ei}r_i + f_e = 0 \qquad \text{1)}$$

$$I_i = w_{ie}r_e - w_{ii}r_i + f_i = 0 \qquad \text{2)}$$

where $w_{xy}$ is the mean coupling strength from neurons in population $y$ to neurons in population $x$. This can be written in matrix form as

$$Wr + f = 0 \qquad \text{3)}$$

where

$$W = \begin{bmatrix} w_{ee} & w_{ei} \\ w_{ie} & w_{ii} \end{bmatrix}, r = \begin{bmatrix} r_e \\ r_i \end{bmatrix}, \text{ and } f = \begin{bmatrix} f_e \\ f_i \end{bmatrix}$$

The firing rates of each population can be solved for if the external drives are in the column space of the coupling matrix $W$.

This becomes an important issue if we arbitrarily divide the homogeneous random network into two pools of excitatory and inhibitory neurons and give each pool a different drive. The system can then be expressed using four populations:

$$\begin{bmatrix} w_{ee} & w_{ei} & w_{ee} & w_{ei} \\ w_{ie} & w_{ii} & w_{ie} & w_{ii} \\ w_{ee} & w_{ei} & w_{ee} & w_{ei} \\ w_{ie} & w_{ii} & w_{ei} & w_{ii} \end{bmatrix} \begin{bmatrix} r_{e1} \\ r_{i1} \\ r_{e2} \\ r_{i2} \end{bmatrix} = - \begin{bmatrix} f_{e1} \\ f_{i1} \\ f_{e2} \\ f_{i2} \end{bmatrix} \qquad \text{4)}$$

The only difference between the two pools are the drives $f$; the intra- and inter-pool coupling is identical. This imposed symmetry (note symmetry is between pools, coupling matrix need not be symmetric) makes the coupling matrix $W$ rank 2. The column space is symmetric between the pools and thus solutions for the rates only exist if $f$ is also symmetric (i.e. $f_{e1} = f_{e2}, f_{i1} = f_{i2}$). Therefore, the homogeneous random network driven by spatially inhomogeneous drive cannot obey the balance conditions *per se*.

We can break the pool symmetry if we assume that intra- and inter-pool connections have different weights such as a network with two mutually inhibiting pools. We assume each pool is identical and replace long-range connections with notation $w_{ie_{LONG}}$. This leads to a system of four equations with four unknowns that obey

$$\begin{bmatrix} w_{ee} & w_{ei} & 0 & 0 \\ w_{ie} & w_{ii} & w_{ie_{LONG}} & 0 \\ 0 & 0 & w_{ee} & w_{ei} \\ w_{ie_{LONG}} & 0 & w_{ie} & w_{ii} \end{bmatrix} \begin{bmatrix} r_{e1} \\ r_{i1} \\ r_{e2} \\ r_{i2} \end{bmatrix} = - \begin{bmatrix} f_{e1} \\ f_{i1} \\ f_{e2} \\ f_{i2} \end{bmatrix} \quad 5)$$

This sparse matrix is full rank, so its columns span all possible input vectors *f*. Solving this system gives expressions for the firing rates of each pool. The expressions for the excitatory and inhibitory rates of pool 1 are

$$r_{e1} = \frac{w_{ie_{LONG}} f_{i2} - \frac{w_{ie_{LONG}} f_{e2} w_{ii}}{w_{ei}} + (f_{i1} - \frac{f_{e1} w_{ii}}{w_{ei}})(\frac{w_{ee} w_{ii}}{w_{ei}} - w_{ie})}{\left(\frac{w_{ee} w_{ii}}{w_{ei}} - w_{ie}\right)^2 - w_{ie_{LONG}}^2} \quad 6)$$

$$r_{i1} = \frac{w_{ie} r_{e1} + w_{ie_{LONG}} r_{e2} + f_{i1}}{w_{ii}} \quad 7)$$

The expressions for pool 2 are the same with indices (1,2) exchanged. Note that when long-range connections are deleted, these expressions reduce to single-pool balanced state equations as expected.

Inputs balance to zero in units of threshold in this theory, but the threshold for an integrate-and-fire neuron in a network with a finite number of neurons is unknown. To account for this, we introduce variable threshold parameters $T_e$ and $T_i$ for excitatory and inhibitory neurons, which we subtract from the feedforward drive. The $W$ parameters in our theory equations can be calculated analytically from our simulations parameters with

$$W_{theory} = W_{sim} \sqrt{k} * \tau_s \quad 8)$$

This comes from the fact each neuron receives exactly *k* synaptic inputs of each connection type (excitatory or inhibitory), but individual synapses are scaled as $\frac{1}{\sqrt{k}}$. We do not normalize synaptic strength by the time constant so the time integral over the synaptic input will scale with the time constant. Alternatively, theory parameters can be calculated empirically by solving the equation

$$I_{ij} = w_{ij} r_j \quad 9)$$

By measuring the synaptic input from population *j* to population *i* and dividing by the firing rate of population *j* to recover theory parameters. We found that both methods return virtually identical results.

We first compared the single-pool balanced-state theory to simulations. We found (see Figure S3) that the predicted spiking rate trends from single-pool balanced state theory were similar to the simulations but spiking rates predicted from theory were systematically higher than we observed in simulations.

**Figure S3**

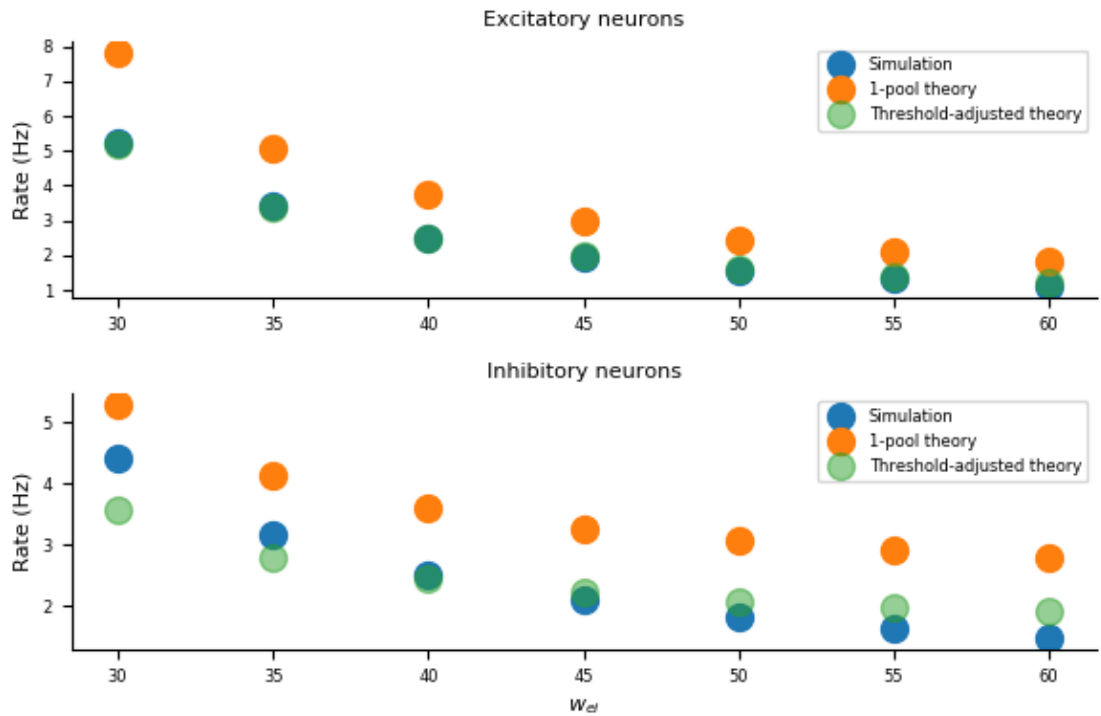

**Figure S3: comparison of single-pool balanced state theory to simulations. A)** Excitatory firing rates in simulations and in theory predictions. **B)** Same as (A), but for inhibitory neurons.

We next tested to see if whether a single value of $T_e$ and $T_i$ can account for the bias in the theory over the range of coupling parameters we tested. We searched this two-dimensional space for 100 values of each threshold parameter, spaced evenly from -3 to 3. For each set of thresholds, we compared theory predictions to observed rates and chose thresholds that minimized the summed Euclidean distance for excitatory and inhibitory neurons. This resulted in predictions that closely matched simulations within a reasonable range of coupling parameters

With the capability to predict firing rates in single-pool networks of neurons, we turned to applying two-pool balanced state theory to mutual inhibition networks. These networks can exhibit WTA dynamics when mutual inhibition strength ($w_{ie_{LONG}}$) is strong enough. In order to apply two-pool theory to networks in the both-active state as well as the WTA state, we compared simulations to theory over a range of mutual inhibition strengths. We used the same process described above of recording rates from simulations, calculating theory parameters, scanning threshold parameters, and comparing theory predictions to our results.

In the both-active regime, two-pool balanced state theory with a threshold adjustment closely matches the firing rates seen in our simulations. As mutual inhibition strength was increased, we observed that the theory equations become singular and the theory breaks down (see Figure S4). The location of the singularity demarked the transition from both-active to WTA as shown in the main text. Specifically, the numerator and denominator of the rate equations become zero simultaneously. To make the singularity visible, we perturbed one of the $f$ terms by a miniscule amount ($0.01 \frac{mV}{sm}$). Below the numerator and denominator components of our equations are plotted to make this clear. In Figure 5 in the main text, this tiny perturbation in inputs was used for both theory and simulation. The same threshold adjustment that was optimal for the 2-pool theory was used in the 1-pool theory.

**Figure S4**

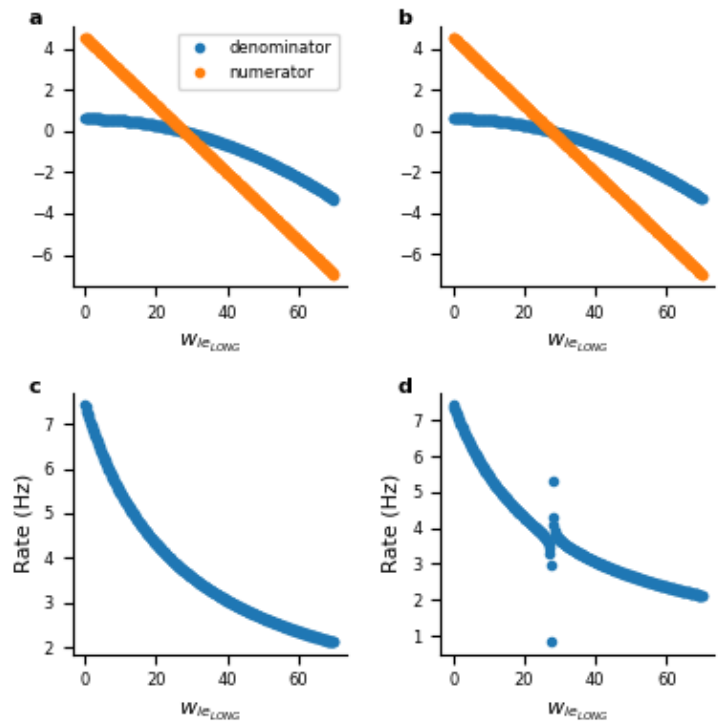

**Figure S4: Visualizing the singularity in the theoretical rate equations.** **A)** Numerator and denominator of rate equations for various mutual inhibition strengths. Both cross 0 simultaneously. **B)** Same as (A), but with $f_{e2}$ modified by 0.01. **C)** Rates predicted by theory. Singularity is not visible. **D)** Perturbing one of the drive terms makes numerator and denominator cross zero at slightly different locations, making the singularity obvious.

## 5. Adaptation effect in the unstructured network

Since random networks amplify asymmetries in their time-dependent feedforward inputs, we tested whether adding a neuronal fatigue process could result in oscillations when the inputs are constant but asymmetric. Drive parameters were $f_{e1} = 3.\frac{mV}{ms}, f_{e2} = 3.5\frac{mV}{ms}, f_{i1} = \frac{2.9mV}{ms}, f_{i2} = 3.4\frac{mV}{ms}$. Coupling parameters were the same as used in main text. Strong adaptation was required to produce oscillations, with an amplitude of $5\frac{mV}{ms}$ and a time constant of 1 second. Oscillations were most pronounced when both excitatory and inhibitory neurons adapted (figure S5). Each excitatory population had oscillating activity levels, but this does not appear to reflect rivalry since each population appears to have an independent oscillatory frequency. This transient behavior disappeared entirely when the adaptation variable was initialized randomly across neurons.

**Figure S5**

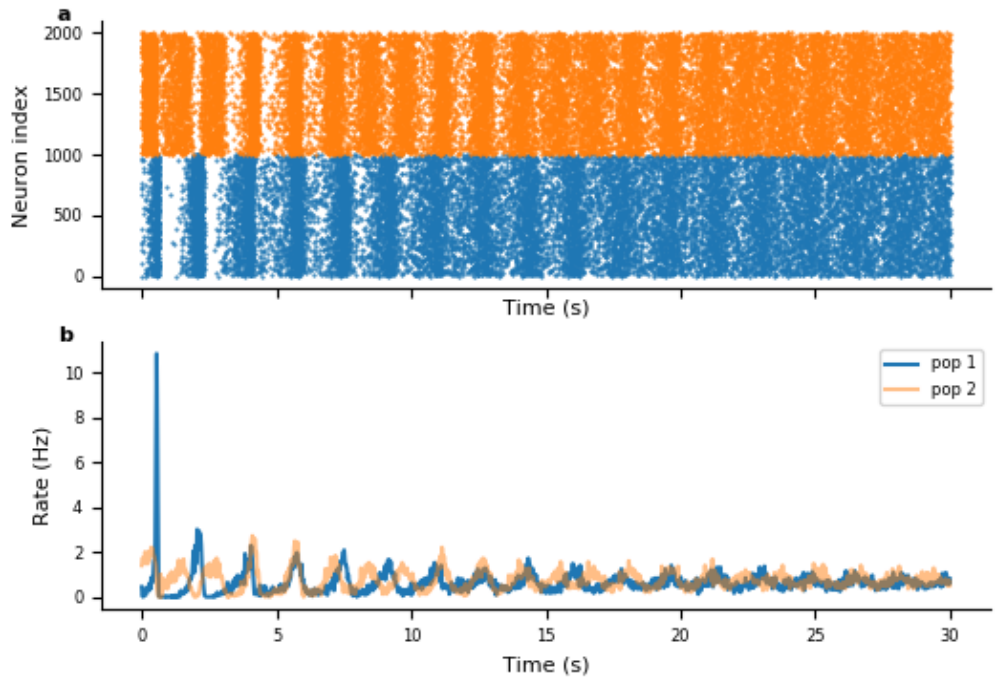

**Figure S5: Decaying oscillations in a random network with constant asymmetric drive and adaptation. A)** Excitatory raster for two excitatory populations with different drive in the same network. **B)** Population averaged firing rates in contiguous 25ms time windows for each excitatory population.